\documentclass[aps,prl,amsmath,amssymb,twocolumn,showpacs,superscriptaddress,notitlepage,longbibliography]{revtex4-2}
\usepackage{bm}
\usepackage{xcolor}
\usepackage{graphicx}
\usepackage{color}
\usepackage{multirow}
\usepackage{graphicx}
\usepackage{subfigure}
\usepackage{dcolumn}
\usepackage{bm}
\usepackage{amsfonts}
\usepackage{mathrsfs}
\usepackage{amssymb}
\usepackage{amsmath}
\usepackage{color}
\usepackage{chngcntr}
\usepackage{stmaryrd}
\usepackage{adjustbox}
\usepackage{microtype}
\definecolor{mycolor}{RGB}{0, 0, 255} 
\usepackage[colorlinks=true,linkcolor=mycolor,anchorcolor=mycolor,citecolor=mycolor,urlcolor=mycolor]{hyperref}
\date{\today}
\allowdisplaybreaks[3]
\usepackage{txfonts}
\usepackage[T1]{fontenc}
\begin{document}
\addtocontents{toc}{\protect\setcounter{tocdepth}{-1}} %
%

\title{Open Quantum Theory of Shot Noise in Dissipative Chiral Transport}

\author{Ming Gong}
\email{minggong@cat.phys.s.u-tokyo.ac.jp}
\email{minggong@pku.edu.cn}
\affiliation{Department of Physics, The University of Tokyo, 7-3-1 Hongo, Tokyo 113-0033, Japan}
\affiliation{International Center for Quantum Materials, School of Physics, Peking University, Beijing 100871, China}
\author{Masahito Ueda}
\email{ueda@cat.phys.s.u-tokyo.ac.jp}
\affiliation{Department of Physics, The University of Tokyo, 7-3-1 Hongo, Tokyo 113-0033, Japan}
\affiliation{Fundamental Quantum Science Program (FQSP), TRIP Headquarters, RIKEN, Wako 351-0198, Japan}
\affiliation{Institute for Physics of Intelligence, The University of Tokyo, 7-3-1 Hongo, Tokyo 113-0033, Japan}

\begin{abstract}
We develop an open quantum theory for shot-noise dynamics in dissipative chiral transport. By mapping a system under consideration onto a quantum circuit, we show that current noise is governed by two competing factors: the average occupancy distribution and particle-number fluctuations. With energy fully relaxed, shot noise is strongly suppressed, reflecting the stacking of electrons into lower energy states due to dissipation. This process quenches the partition noise from partially occupied levels, and finally isolates the residual noise protected by strong $U(1)$ symmetry. Moreover, selectively heating the source against the bath uncovers the underlying competition between the noise contributions from the occupancy distribution and those from the particle-number fluctuations. It triggers a sign reversal in inter-channel correlation noise, a signature masked by seemingly identical single-channel thermal noises. We propose an inversion scheme to experimentally reconstruct the hidden occupancy distribution directly from measurable noise cumulants.
\end{abstract}

\maketitle

\allowdisplaybreaks
\emph{Introduction}---As a conductor scales up from the mesoscopic to the macroscopic regime, shot noise tends to be suppressed, leaving only the Johnson-Nyquist noise as the dominant fluctuations in electronic transport~\cite{johnson_thermal_1928,nyquist_thermal_1928,brillouin_fluctuations_1934,blanter_shot_2000}. While early studies show that pure dephasing cannot fully suppress shot noise, energy dissipation is widely recognized as the main mechanism for this suppression~\cite{beenakker_semiclassical_1991,shimizu_effects_1992,beenakker_suppression_1992,liu_suppression_1994,liu_nyquist_1994,de_jong_semiclassical_1995,steinbach_observation_1996,nazarov_quantum_2003,pala_effect_2004,kumada_shot_2015,beenakker_pure_2026}. A microscopic understanding based on open quantum theory and the key variables governing the noise dynamics remain elusive.
To bridge this gap, multi-channel chiral transport systems offer an ideal theoretical platform. Apart from serving as building blocks for dissipationless electronics and topological quantum computation~\cite{chang_experimental_2013,kou_scale-invariant_2014,chang_zero-field_2015,yasuda_quantized_2017,marguerite_imaging_2019,zhao_tuning_2020,chang_colloquium_2023,li_emergent_2024,yan_rules_2024,fijalkowski_balanced_2024,qi_chiral_2010,beenakker_search_2013,sato_topological_2017,lian_topological_2018}, chiral transport systems naturally bypass the complexities of backscattering~\cite{buttiker_absence_1988,beenakker_random-matrix_1997,kumada_shot_2015,ma_graphene_2018,beenakker_monitored_2025,beenakker_shot_2025,beenakker_pure_2026}, isolating the essential physics of noise suppression through dissipation.

In this Letter, we develop an open quantum theory for an $N$-channel chiral transport system by mapping it onto a multi-layer fermionic quantum circuit in energy space. This natural microscopic framework overcomes the difficulty of traditional Green's functions in treating correlations, dissipation, and current noise simultaneously~\cite{datta_electronic_1995,haug_quantum_2008,nazarov_quantum_2009}. Leveraging an effective cumulant generating function (CGF), we show that the dissipative noise dynamics is governed by the interplay between two distinct contributions. The first contribution relates to the average occupancy distribution $\langle M_k \rangle$, defined as the average number of energy levels occupied by $k$ particles. The second contribution is proportional to the fluctuation of the total particle number $\langle\Delta N_{\rm tot}^2\rangle$. Specifically, for the first part, the number of partially occupied levels $M_k$ ($k=1,\dots,N-1$) gives rise to the partition-induced fluctuations. It dominates the single-channel noise $S_{11}$ and produces a negative inter-channel correlation noise $S_{12}$, reminiscent of the fermionic Hanbury Brown and Twiss effect~\cite{henny_fermionic_1999,oliver_hanbury_1999}. As dissipation forces electrons to pack into lower energy states, $S_{11}$ and $S_{12}$ are suppressed due to the quenching of partial occupancies. The second part characterizes the initial particle-number fluctuations of injected electrons, which are protected by the strong $U(1)$ symmetry of particle conservation~\cite{buca_note_2012,albert_symmetries_2014}. When the system fully relaxes to a cold bath, the partition-induced contribution completely vanishes. The conserved initial fluctuations then survive to provide a residual noise, which leads to a positive $S_{12}$. On the basis of this picture, we find that tuning the system from the bath-heating to the source-heating regime impacts the noise distinctly. $S_{12}$ undergoes a sign reversal from negative to positive, while $S_{11}$ remains seemingly invariant. 
This demonstrates that identical macroscopic thermal noise originates from fundamentally distinct microscopic mechanisms.

Verifying this microscopic mechanism experimentally requires extracting both $\langle\Delta N_{\rm tot}^2\rangle$ and $\langle M_k \rangle$. While the former is easily obtained from total fluctuations, the latter remains hidden from the standard second-order noise. Going beyond the second order, our effective CGF yields an inversion scheme, showing that measuring noise cumulants up to the $N$-th order in an $N$-channel system can reconstruct $\langle M_k \rangle$. We numerically validate this inversion framework in an $N=3$ system. The quantitative agreement with matrix product state (MPS) simulations demonstrates its applicability in experimental setups, enabling the extraction of the key microscopic information governing the noise dynamics.
\begin{figure}[htbp]
	\centering
	\includegraphics[width=0.48\textwidth]{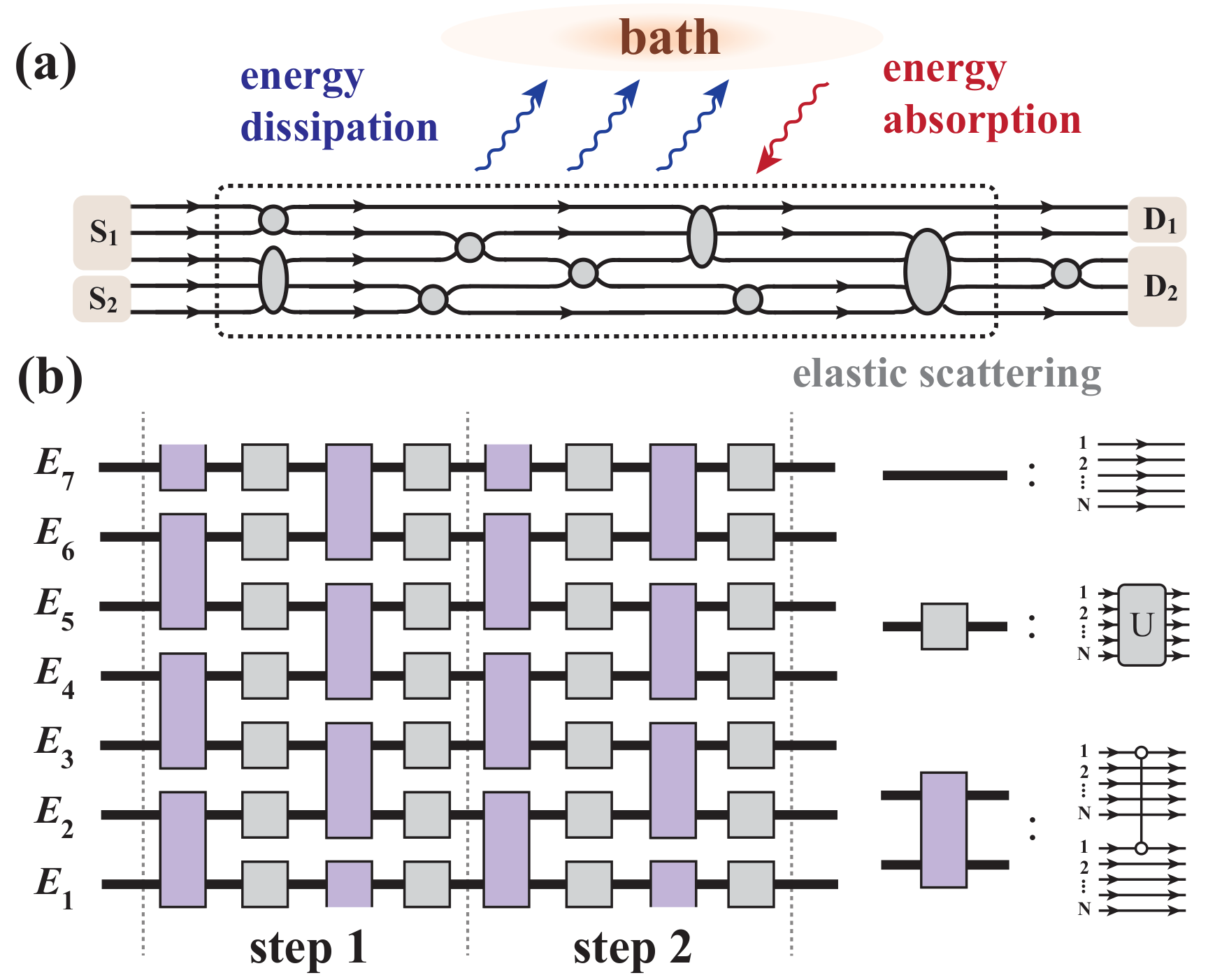}
	\caption{Illustration of the multi-channel chiral transport system and its mapping to a quantum circuit. (a) Multi-channel dissipative chiral transport system. Black lines with arrows denote chiral channels from sources (${\rm S}_{1,2}$) to drains (${\rm D}_{1,2}$). Electrons undergo inter-channel elastic scattering alongside energy exchange with the bath.  (b) Layered quantum circuit mapping. The circuit is structured into discrete energy layers (with energy levels ranging from $E_1$ to $E_7$). Thick lines represent $N$ transport channels at each level. The evolution proceeds in discrete steps. Grey boxes represent intra-layer random unitary gates $\hat U^{(m)}$. Purple boxes represent Kraus operators $\hat{K}_{\uparrow}$, $\hat{K}_{\downarrow}$, and $\hat{K}_{0}$.}
	\label{fig1}
\end{figure}

\emph{Model}---We consider a chiral transport system with $N$ parallel channels, as shown in Fig.~\ref{fig1}(a). Electrons are injected from source terminals ${\rm S}_1, {\rm S}_2, \dots$. During transport, they undergo disorder-induced elastic scattering and exchange energy with the bath before reaching the drain terminals ${\rm D}_1, {\rm D}_2, \dots$. We assume all drains are grounded at zero temperature. In principle, each terminal can couple to multiple channels [as depicted in Fig.~\ref{fig1}(a) for source or drain terminals]; for clarity here we assume each terminal only couples to one channel. 

We map this transport process onto a fermionic quantum circuit, as depicted in Fig.~\ref{fig1}(b). In typical setups such as quantum Hall edge states or graphene $p$--$n$ junctions~\cite{abanin_quantized_2007,williams_quantum_2007,kumada_shot_2015}, an applied bias $V$ opens an active energy window $eV$ that accommodates $M = eV/\Delta E$ discrete energy levels, where $\Delta E = h v_F/l$ is the intrinsic level spacing determined by the transport distance $l$ and the Fermi velocity $v_F$~\cite{datta_electronic_1995,nazarov_quantum_2009}. During a characteristic transit time $\Delta t = h/\Delta E$, exactly one electron is injected into each level. This naturally justifies discretizing the relevant energy space into $M$ equally spaced levels ($E_{m+1}-E_{m}=\Delta E$) with $N$ chiral channels included in each level $E_m$. The system is thus fully described by the annihilation operators $\hat c_{m,n}$ ($m=1,\dots,M$; $n=1,\dots,N$). At finite temperature, the density matrix of injected electrons then reads
\begin{equation}
\label{inject_density_matrix}
\hat\rho_{\rm in} = \prod_{m,n}\mathcal{Z}_{m,n}^{-1} \exp \left[ -(E_m - \mu_{{\rm in}, n}) \hat{n}_{m,n} / k_{B}T_{{\rm in}, n} \right].
\end{equation}
Here, $\hat{n}_{m,n} = \hat{c}_{m,n}^\dagger \hat{c}_{m,n}$ is the particle-number operator, while $T_{{\rm in}, n}$ and $\mu_{{\rm in}, n}$ denote the temperature and chemical potential of the injected electrons in the $n$-th channel (terminal). The normalization factor is $\mathcal{Z}_{m,n} = 1 + \exp[-(E_m - \mu_{{\rm in}, n})/k_{B}T_{{\rm in}, n}]$. In the following we set $k_{B}=1$ for simplicity.

The elastic scattering is modeled by intra-layer unitary gates. The many-body scattering operator for energy level $m$ is
\begin{equation}
	\hat U^{(m)} = \exp \left[ \sum_{i,j=1}^N (\ln s^{(m)})_{ij} \hat c_{m,i}^\dagger \hat c_{m,j} \right].
	\label{Eq_haar_unitary}
\end{equation}
Here, $s^{(m)}$ is a single-particle Haar-random $U(N)$ scattering matrix, chosen independently for each layer [see gray boxes in Fig.~\ref{fig1}(b)]. This means electrons are fully mixed across all channels at each level. As conductors scale up to the macroscopic regime, this assumption captures the chaotic nature of the scattering process in the strong-disorder and long-distance transport regime~\cite{beenakker_random-matrix_1997,oberholzer_shot_2002,li_disorder_2008,beenakker_monitored_2025,beenakker_shot_2025,beenakker_pure_2026}.

To model heat exchange with a bath at temperature $T_{\rm bath}$, we introduce inter-layer jump operators $\hat{K}_\alpha$ ($\alpha \in \{\uparrow, \downarrow, 0\}$) corresponding to energy absorption, dissipation, and no energy exchange between adjacent layers [see purple boxes in Fig.~\ref{fig1}(b)]. These operators are constructed to satisfy the detailed balance condition $\gamma_\uparrow = \gamma_\downarrow \exp(-\Delta E/T_{\rm bath})$, where $\gamma_\downarrow \equiv \gamma_0$ defines the intrinsic dissipation rate. Restricting jumps to adjacent layers is a minimal assumption that captures the fundamental physics of dissipation. Our theory shows that the key variables that determine noise dynamics are independent of the jump range. Crucially, both the scattering gates $\hat{U}^{(m)}$ and the dissipative jump operators $\hat{K}_{\alpha}$ commute with the total particle-number operator $\hat{N}_{\rm tot}=\sum_{m,n}\hat{n}_{m,n}$, strictly enforcing a strong $U(1)$ symmetry in the open quantum dynamics~\cite{buca_note_2012,albert_symmetries_2014}. The exact matrix representations of $\hat{K}_\alpha$ are provided in the Supplemental Materials~\cite{note_supp}.

The dynamical evolution under scattering and dissipation is modeled using a brick-wall quantum circuit. A single evolution step $\tau$, separated by the vertical dashed lines as depicted in Fig.~\ref{fig1}(b), consists of a four-substep sequence of alternating intra-layer scattering gates and inter-layer dissipative jump operators. This sequence defines a completely positive trace-preserving (CPTP) superoperator $\mathcal{E}_\tau$~\cite{breuer_theory_2007,nielsen_quantum_2010}, governing the discrete-time master equation $\hat\rho(\tau) = \mathcal{E}_\tau[\hat\rho(\tau-1)]$. The rigorous step-by-step construction of $\mathcal{E}_\tau$ is detailed in the Supplemental Materials~\cite{note_supp}. Consequently, an $L$-step evolution maps the initial state to
\begin{equation}
\hat\rho(L) = \mathcal{E}_L\circ \mathcal{E}_{L-1}\circ \cdots \circ\mathcal{E}_1 [\hat\rho_{\rm in}].
\end{equation}

In the following numerical simulations, we unravel the discrete-time master equation into stochastic quantum trajectories, where the pure states are represented as MPS~\cite{verstraete_quantum_2009,nielsen_quantum_2010,schollwock_density-matrix_2011,paeckel_time-evolution_2019,weimer_simulation_2021,xiang_density_2023,note_supp,yadalam_process-tensor_2026,itensor}.

\emph{Occupancy distribution and effective cumulant generating function}---Shot noise originates from the particle-number fluctuations of arriving electrons. This behavior is captured by the framework of full counting statistics (FCS)~\cite{levitov1993charge,nazarov_circuit_2002,pilgram_full-counting_2006,forster_voltage_2007}. The CGF $\mathcal{F}$ obtained from the moment generating function (MGF) $\mathcal{G}$ for the FCS of transmitted particles reads
\begin{equation}
\mathcal{F}(\boldsymbol{\lambda}) = \ln \mathcal{G}(\boldsymbol{\lambda}),\quad \mathcal{G}(\boldsymbol{\lambda})=\left\langle e^ {\sum_i \lambda_i \hat{N}_i }\right\rangle.
	\label{CGF_original}
\end{equation}
Here, $\boldsymbol{\lambda}:=(\lambda_1,\lambda_2,\dots,\lambda_N)$, represents the counting field~\footnote{In contrast to the conventional characteristic function with a complex phase $e^{i\lambda \hat{N}}$ widely used in quantum transport, we define the generating function using a real parameter $\lambda$ for algebraic simplicity. Both formalisms are equivalent for extracting the cumulants of the fluctuations.}, $\hat{N}_i = \sum_{m=1}^M \hat{c}_{m,i}^\dagger \hat{c}_{m,i}$ is the particle-number operator for the $i$-th channel across all energy levels, and the angle brackets $\langle \cdots \rangle={\rm Tr}[\cdots \hat\rho(L)]$ denotes the ensemble average under a quenched scattering configuration, which is equivalently evaluated by averaging over an ensemble of unraveled quantum trajectories. The cumulants are extracted by evaluating the derivatives of $\mathcal{F}$ with respect to $\boldsymbol{\lambda}$ at $\boldsymbol{\lambda}=0$. For instance, the mean particle-number in channel $i$ is given by the first derivative, $\langle N_i \rangle = \partial_{\lambda_i} \mathcal{F} |_{\boldsymbol{\lambda}=0}$. Regarding the second-order noise properties, the single-channel fluctuation in channel 1 is determined by $\langle \Delta N_1^2 \rangle = \partial_{\lambda_1}^2 \mathcal{F} |_{\boldsymbol{\lambda}=0}$, where $\Delta N_1 = N_1 - \langle N_1 \rangle$. Similarly, the correlated fluctuation between channels 1 and 2 is  $\langle \Delta N_1 \Delta N_2 \rangle = \partial_{\lambda_1} \partial_{\lambda_2} \mathcal{F} |_{\boldsymbol{\lambda}=0}$. 
\begin{figure}[t]
	\centering
	\includegraphics[width=0.485\textwidth]{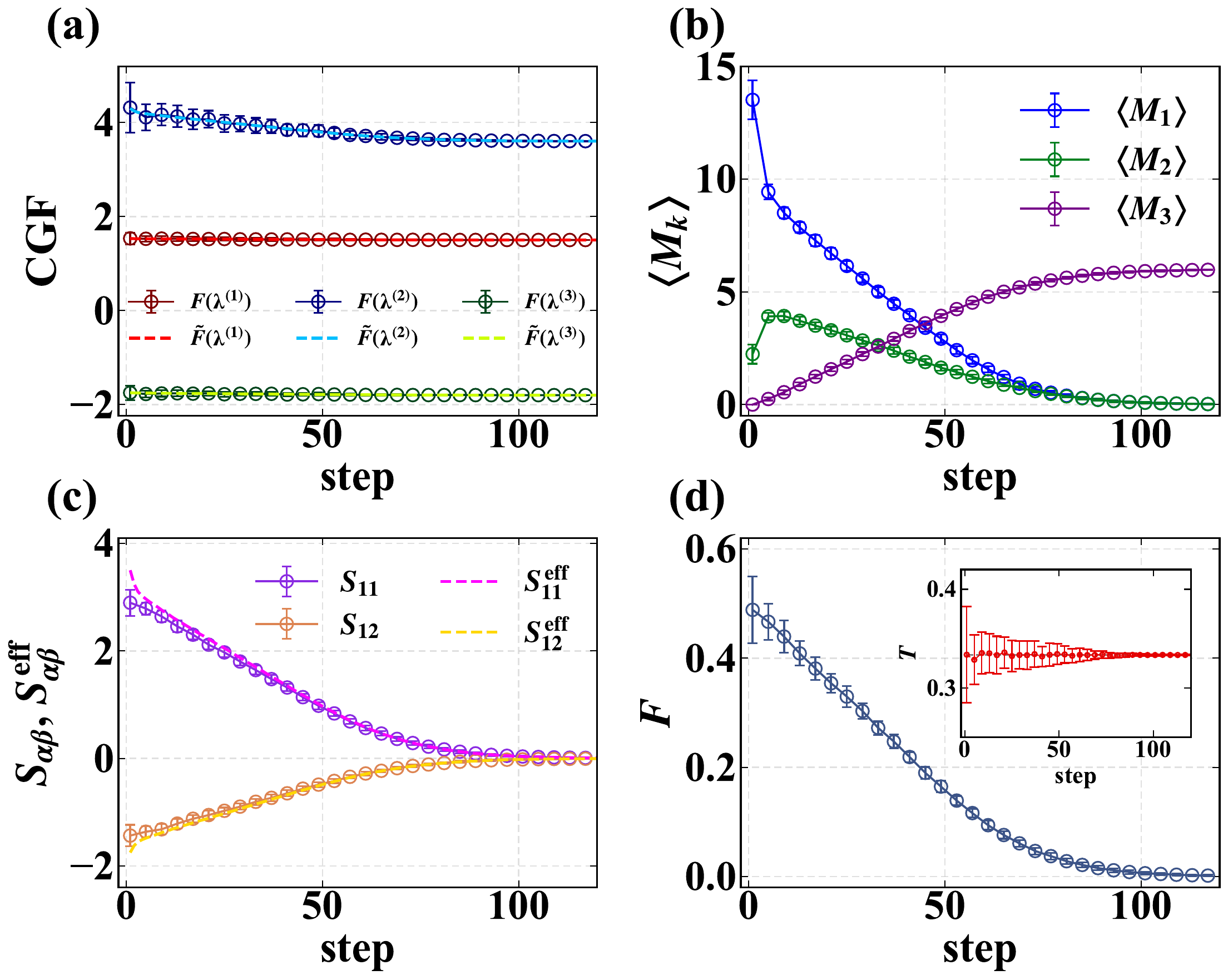}
	\caption{Cumulant generating function (CGF) and shot noise suppression in an $N=3$ chiral transport system. The system parameters are set to $\gamma_0=0.7$, $\mu_{\rm in}=18 \Delta E$, and $T_{\rm in}=T_{\rm bath}=0$. (a) Numerical microscopic CGF $\mathcal{F}(\boldsymbol{\lambda})$ and effective CGF $\tilde{\mathcal{F}}(\boldsymbol{\lambda})$, evaluated at three sets of values: $\boldsymbol{\lambda}^{(1)}=[0.15, 0.10, 0.00]$, $\boldsymbol{\lambda}^{(2)}=[0.60, -0.10, 0.10]$, and $\boldsymbol{\lambda}^{(3)}=[-0.10, -0.20, 0.00]$. (b) Average occupancy distribution $\langle M_k \rangle$ ($k=1,2,3$). (c) $S_{11}$ and $S_{12}$, compared with the effective noise $S_{11}^{\rm eff}$ and $S_{12}^{\rm eff}$ (in units of $2e^2/\Delta t$). (d) Evolution of the Fano factor $F$. Inset: Transmission coefficient $T$. Error bars indicate variations across random quenched scattering configurations.}
	\label{fig2}
\end{figure}

Solving $\mathcal{F}(\boldsymbol{\lambda})$ exactly is analytically intractable. Nevertheless, the Haar-random scattering provides a coarse-graining scheme to drastically reduce the complexity in the analysis. Notably, the particle-number occupancy in each level, $\mathbf{n} := (n_1, \dots, n_M)$, remains well-defined at any step along a quantum trajectory, and the random scattering in Eq.~(\ref{Eq_haar_unitary}) effectively mixes the intra-level channels and strongly suppresses inter-level correlations. Therefore, the occupancy distribution $\mathbf{M} := (M_0, M_1, \dots, M_N)$, defined as $M_i \equiv \sum_{m=1}^M \delta_{n_m, i}$, governs the statistical behavior of quantum trajectories. Consequently, the physics of $\mathcal{F}(\boldsymbol{\lambda})$ is captured by $\mathbf{M}$. We thus propose an effective CGF $\tilde{\mathcal{F}}$ and an effective MGF $\tilde{\mathcal{G}}$ expressed as
\begin{subequations}
\label{eq:effective_CGF_all}
\begin{align}
\tilde{\mathcal{F}}(\boldsymbol{\lambda}) &= \ln \tilde{\mathcal{G}}(\boldsymbol{\lambda}),\quad \tilde{\mathcal{G}}(\boldsymbol{\lambda})=\langle e^{\Psi(\boldsymbol{\lambda})} \rangle_{\mathbf{M}}, \label{eq:effective_CGF_a} \\
\Psi(\boldsymbol{\lambda}) &= \sum_{k=1}^N M_k \ln \left[\sum_{i_1 < \dots < i_k} e^{\lambda_{i_1} + \cdots + \lambda_{i_k}}/\binom{N}{k}   \right]. \label{eq:effective_CGF_b}
\end{align}
\end{subequations}
Here, $\langle \cdots \rangle_{\mathbf{M}}$ denote the statistical average over $\mathbf{M}$, with $\binom{N}{k}$ the binomial coefficient. In fact, $e^{\Psi(\boldsymbol{\lambda})}$ represents the conditional MGF for a fixed $\mathbf{M}$, obtained under the approximations of perfect intra-level channel mixing and inter-level statistical independence. 

We numerically simulate a chiral transport system with $N=3$ (a configuration we adopt for the remainder of this Letter). By averaging over unraveled quantum trajectories, we directly compute the exact CGF $\mathcal{F}$ [Eq.~(\ref{CGF_original})]. Simultaneously, we extract the distribution $\mathbf{M}$ from each trajectory and average Eqs.~(\ref{eq:effective_CGF_a}) and (\ref{eq:effective_CGF_b}) over the ensemble to obtain the effective CGF $\tilde{\mathcal{F}}$. As shown in Fig.~\ref{fig2}(a), the excellent agreement between $\mathcal{F}$ and $\tilde{\mathcal{F}}$ demonstrates that our proposed effective CGF accurately captures the essential features of the exact CGF. This compact expression of $\tilde{\mathcal{F}}$ reveals that the underlying noise dynamics is intrinsically encoded within the statistics of $\mathbf{M}$. It serves as the key to the physical picture of noise suppression.
 
\emph{Picture of shot noise suppression}---We consider a specific transport setup. Suppose terminal 1 is biased with a voltage $V$ and all other terminals are grounded. We set a uniform incident temperature across all channels, i.e., $T_{\rm in,1}=T_{\rm in,2}=T_{\rm in,3} \equiv T_{\rm in}$ in Eq.~(\ref{inject_density_matrix}). Focusing first on the zero-temperature limit ($T_{\rm in}=T_{\rm bath}=0$), during a typical transit time $\Delta t$, the average current received at drain terminal $\alpha$ is $\langle I_\alpha \rangle = (e/\Delta t) \langle N_\alpha \rangle$. This naturally defines the transmission coefficient $\mathcal{T}_\alpha = \langle I_\alpha \rangle / I_{\rm in} = \langle N_\alpha \rangle / \langle N_{\rm tot} \rangle$, normalized by the total incident current $I_{\rm in} = (e/\Delta t) \langle N_{\rm tot} \rangle$. For a three-channel system, the transmission coefficient to a specific channel approaches $1/3$, as shown in the inset of Fig.~\ref{fig2}(d). To characterize the fluctuations, the zero-frequency noise spectral density between channels $\alpha$ and $\beta$ is evaluated as $S_{\alpha\beta} = (2e^2/\Delta t) \langle \Delta N_\alpha \Delta N_\beta \rangle$~\cite{nazarov_quantum_2009}. Consequently, the Fano factor for channel 1, given by $F = S_{11} / (2e \langle I_1 \rangle) = \langle \Delta N_1^2 \rangle / \langle N_1 \rangle$, serves as a dimensionless measure of the relative noise strength.

Our effective CGF $\tilde{\mathcal{F}}$ provides a transparent framework to extract the dominant noise contributions. Differentiating $\tilde{\mathcal{F}}$ [Eq.~(\ref{eq:effective_CGF_a})] yields the effective noise $S_{\alpha\beta}^{\rm eff}$ given by
\begin{subequations}
\label{eq:noise_spectra}
\begin{align}
S_{11}^{\rm eff} &\propto \frac{\partial^2 \tilde{\mathcal{F}}}{\partial \lambda_1^2} \bigg|_{\boldsymbol{\lambda}=0} = \sum_{k=1}^N  \frac{k(N-k)}{N^2}\langle M_k \rangle + \frac{1}{N^2} \langle\Delta N_{\rm tot}^2\rangle, \label{eq:S11} \\
S_{12}^{\rm eff} &\propto \frac{\partial^2 \tilde{\mathcal{F}}}{\partial \lambda_1 \partial \lambda_2} \bigg|_{\boldsymbol{\lambda}=0} = - \sum_{k=1}^N  \frac{k(N-k)}{N^2(N-1)}\langle M_k \rangle + \frac{1}{N^2} \langle\Delta N_{\rm tot}^2\rangle. \label{eq:S12}
\end{align}
\end{subequations}
At zero temperature, the conservation of $N_{\rm tot}$ in each quantum trajectory implies that $\langle\Delta N_{\rm tot}^2\rangle=0$. The noise is thus solely governed by the average number of partially filled levels $\langle M_k \rangle$ ($k=1,2,\dots,N-1$). Numerical results show that dissipation significantly alters this distribution. By driving electrons to relax into lower energy states, dissipation induces a densely packed configuration in which fully occupied $\langle M_N \rangle$ and empty $\langle M_0 \rangle$ states dominate, while the partially filled levels $\langle M_k \rangle$ ($k=1,\dots,N-1$) decay to zero [see Fig.~\ref{fig2}(b)]. As a result, $S_{11}$, $S_{12}$ [Fig.~\ref{fig2}(c)], and the Fano factor $F$ [Fig.~\ref{fig2}(d)] are simultaneously suppressed. An excellent agreement between $S_{\alpha\beta}$ and $S_{\alpha\beta}^{\rm eff}$ demonstrates that Eqs.~(\ref{eq:S11}) and (\ref{eq:S12}) successfully capture the essential physics: scattering mixes the channels to generate partition noise, while dissipation stacks electrons into fully occupied states, ultimately quenching the shot noise.

\begin{figure}[t]
	\centering
	\includegraphics[width=0.485\textwidth]{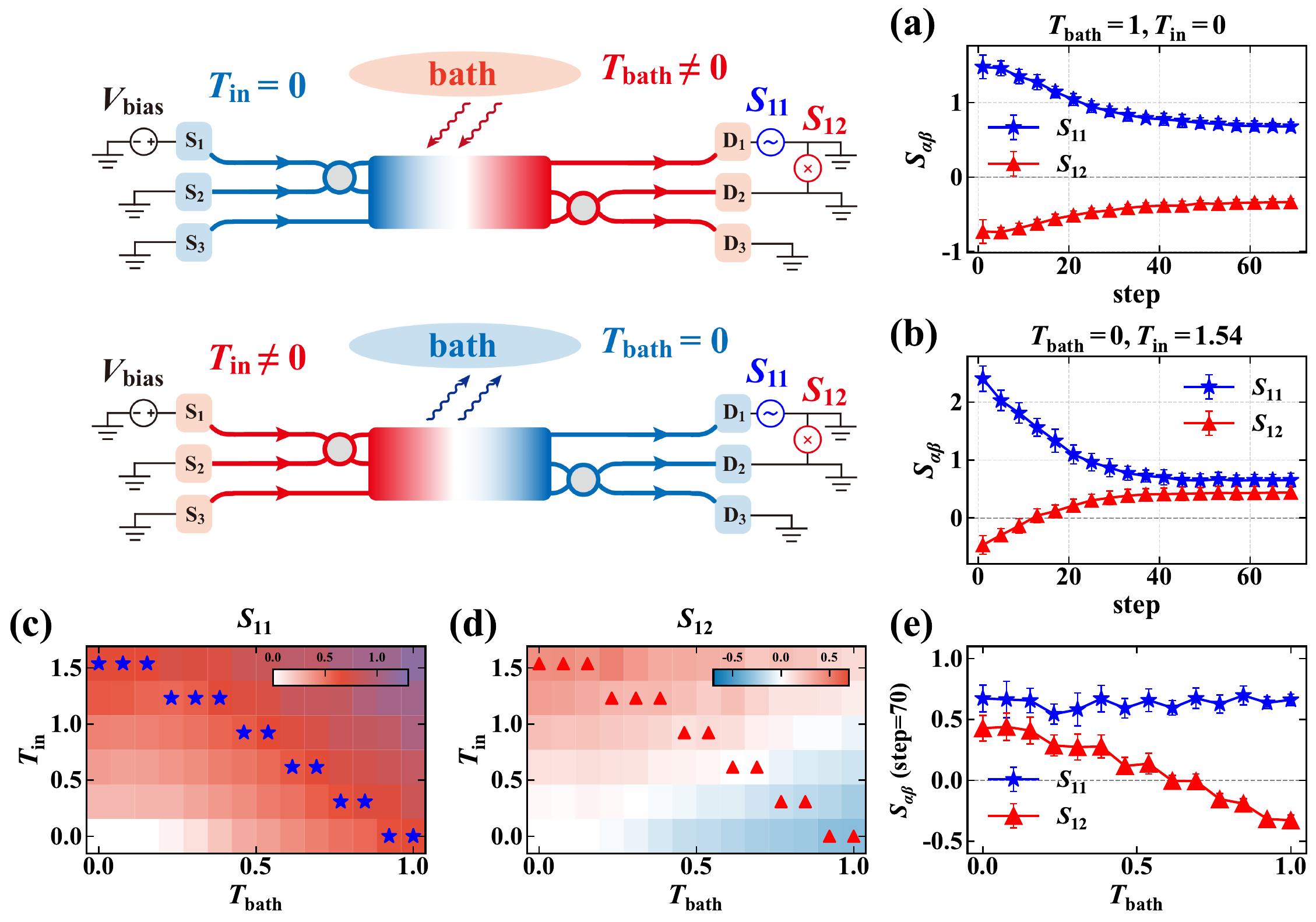}
	\caption{Impacts of the origins of thermal fluctuations on noise dynamics. The schematic illustration shows transport processes dominated by pure bath-heating ($T_{\rm in}=0$) and pure source-heating ($T_{\rm bath}=0$). The system parameters are set to $\gamma_0=0.99$ and $\mu_{\rm in}=9 \Delta E$. (a,b) Evolution of the single-channel noise $S_{11}$ and inter-channel correlation $S_{12}$ for these two cases, corresponding to the rightmost ($T_{\rm bath}=1, T_{\rm in}=0$) and leftmost ($T_{\rm bath}=0, T_{\rm in}=1.54$) marked points in (c) and (d), respectively. (c,d) Steady-state noise $S_{11}$ and $S_{12}$ (at step=70, in units of $2e^2/\Delta t$) as functions of $T_{\rm bath}$ and $T_{\rm in}$. The markers indicate a parameter set where $S_{11}$ is approximately constant. (e) Steady-state noise evaluated along the marked set. Error bars indicate variations across random quenched scattering configurations.}
	\label{fig3}
\end{figure}
\emph{Impact of thermal fluctuations}---We now focus on the distinct noise contributions from $\langle M_k \rangle$ and $\langle\Delta N_{\rm tot}^2\rangle$ induced by thermal fluctuations. First, consider the pure bath-heating limit ($T_{\rm in}=0, T_{\rm bath}>0$), as illustrated in the upper schematic of Fig.~\ref{fig3}. In this case, $\langle\Delta N_{\rm tot}^2\rangle=0$, and the fluctuations are solely governed by the $\langle M_k \rangle$ terms in Eqs.~(\ref{eq:S11}) and (\ref{eq:S12}). Consequently, $S_{12}^{\rm eff}$ is strictly negative, as confirmed numerically in Fig.~\ref{fig3}(a). This reflects the inherent fermionic anti-correlation driven by scattering-induced partitioning, a phenomenon that has been experimentally verified in fermionic Hanbury Brown and Twiss setups~\cite{henny_fermionic_1999,oliver_hanbury_1999}.

In the pure source-heating regime ($T_{\rm in}>0, T_{\rm bath}=0$) (see lower schematic of Fig.~\ref{fig3}), the hot source injects a fluctuating total number of particles. In contact with the cold bath, the system fully relaxes into a compact, fully stacked Fermi state in each quantum trajectory, completely quenching the partition noise (i.e., driving $\langle M_k \rangle \to 0$). However, protected by the strong $U(1)$ symmetry associated with $N_{\rm tot}$ conservation, the source fluctuation dominates, leaving a residue noise proportional to $\langle\Delta N_{\rm tot}^2\rangle$. As a result, $S_{12}^{\rm eff}$ turns positive [Fig.~\ref{fig3}(b)], demonstrating that the partition-induced anti-correlation is overwhelmed by the synchronized fluctuations across channels associated with the fully stacked state. Going beyond phenomenological understandings of relevant experiments based on effective incoherent scattering~\cite{texier_effect_2000,oberholzer_positive_2006,ota_negative_2017}, Eqs.~(\ref{eq:S11}) and (\ref{eq:S12}) shows that this sign reversal is microscopically driven by the competition between partition-induced anti-correlations and strong $U(1)$ symmetry-protected synchronized fluctuations

As a direct consequence of Eqs.~(\ref{eq:S11}) and (\ref{eq:S12}), simultaneously tuning $T_{\rm in}$ and $T_{\rm bath}$ allows $S_{11}^{\rm eff}$ to be maintained nearly constant while $S_{12}^{\rm eff}$ undergoes a sharp sign reversal. To verify this, we extract a representative parameter contour from Figs.~\ref{fig3}(c) and \ref{fig3}(d) where $S_{11}$ is approximately invariant. As the parameters evolve along this contour from the pure bath-heating ($T_{\rm bath}=1, T_{\rm in}=0$) to the pure source-heating ($T_{\rm bath}=0, T_{\rm in}=1.54$), $S_{12}$ transitions from negative to positive [Fig.~\ref{fig3}(e)]. This crossover from anti-correlation to synchronized fluctuations reveals that seemingly identical thermal noise can originate from entirely distinct physical mechanisms.

\begin{figure}[b]
	\centering
	\includegraphics[width=0.485\textwidth]{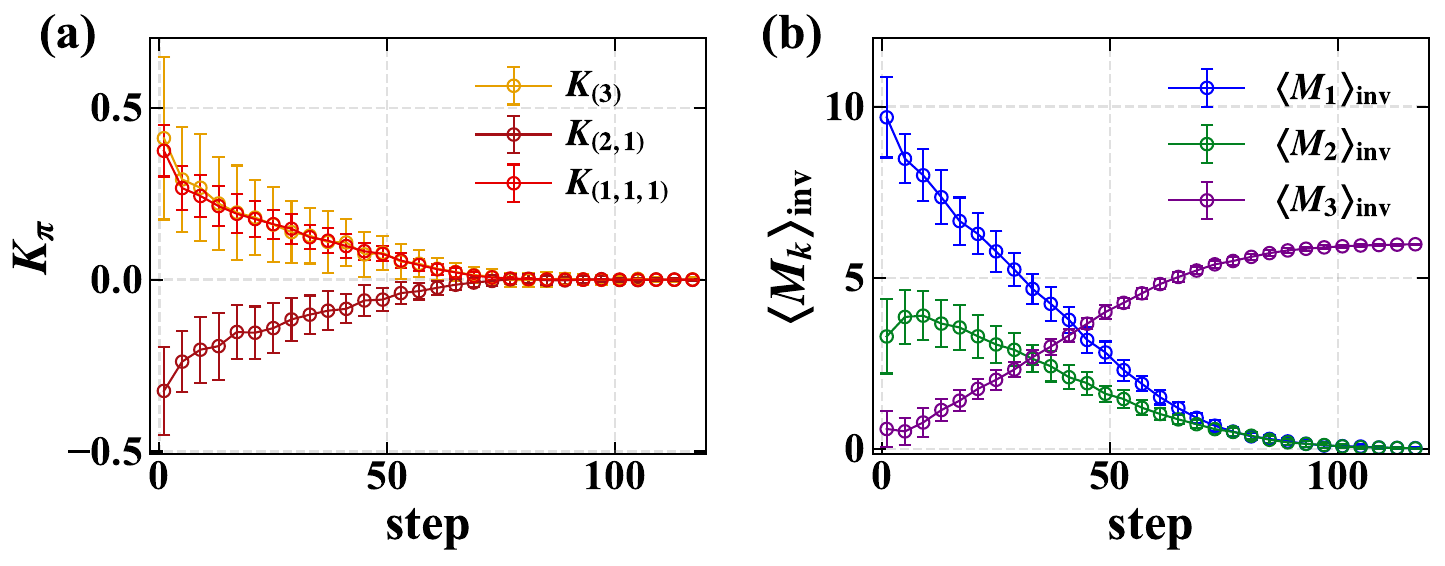}
	\caption{Reconstruction of the occupancy distribution from noise cumulants. All parameters and the averaging over quenched scattering configurations are identical to those in Fig.~\ref{fig2}. (a) Third-order noise cumulants $K_{\pi}$ for different channel partitions $\pi \in \{(3), (2,1), (1,1,1)\}$. (b) Inverted average occupancy distribution $\langle M_k \rangle_{\rm inv}$ ($k=1,2,3$) reconstructed via the inversion scheme, compared with Fig.~\ref{fig2}(b). Error bars indicate variations across random quenched scattering configurations.}
	\label{fig4}
\end{figure}
\emph{Inversion scheme for occupancy distribution}---Since $\langle M_k \rangle$ and $\langle\Delta N_{\rm tot}^2\rangle$ determine the underlying noise dynamics, a fundamental question arises: can we extract them from experimental data? Extracting $\langle\Delta N_{\rm tot}^2\rangle$ is straightforward, as it relies solely on the total input or output fluctuations. However, since the occupancy configuration lacks a direct correspondence to macroscopic observables, extracting $\langle M_k \rangle$ remains highly nontrivial.

We use the effective CGF to establish a theoretical inversion scheme to reconstruct $\langle M_k \rangle$. For a system with $N$ chiral channels, measuring the cumulants of current noise up to the $N$-th order yields a theoretical estimate of $\langle M_k \rangle_{\rm inv}$. For arbitrary $N$, the generalized formula is provided in the End Matter, with its detailed derivation given in the Supplemental Materials~\cite{note_supp}. As the experimental extraction of low-frequency third-order current statistics has been demonstrated in tunnel junctions, quantum point contacts, and diffusive conductors~\cite{bomze_measurement_2005,gershon_detection_2008,pinsolle_non-gaussian_2018}, we present the explicit inversion scheme for $N=3$ for practical implementation, given by
\begin{subequations}
	\label{eq:inversion_N3}
	\begin{align}
		\langle M_1 \rangle_{\rm inv} &= \frac{3}{2}( K_{(2)} - K_{(1,1)} + K_{(3)} - 3K_{(2,1)} + 2K_{(1,1,1)}), \label{eq:M1} \\
		\langle M_2 \rangle_{\rm inv} &= \frac{3}{2}( K_{(2)} - K_{(1,1)} - K_{(3)} + 3K_{(2,1)} - 2K_{(1,1,1)}), \label{eq:M2} \\
		\langle M_3 \rangle_{\rm inv} &= K_{(1)} - \frac{3}{2}(K_{(2)} - K_{(1,1)}) + \frac{1}{2}(K_{(3)} - 3K_{(2,1)} + 2K_{(1,1,1)}). \label{eq:M3}
	\end{align}
\end{subequations}
Here, $K_\pi = \partial_\pi {\mathcal F}(\boldsymbol{\lambda})|_{\boldsymbol{\lambda}=0}$ represents the cumulants evaluated from the microscopic CGF ${\mathcal F}$, where $\pi$ denotes the specific combination of partial derivatives with respect to the counting fields $\lambda_i$. For instance, $\pi=(1,1)$ corresponds to $\partial^2 / (\partial \lambda_1 \partial \lambda_2)$, yielding the covariance $K_{(1,1)} = {\rm Cov}(N_1, N_2)$. Furthermore, $\langle M_0 \rangle_{\rm inv}$ is straightforwardly determined through the constraint $\sum_{k=0}^N \langle M_k \rangle_{\rm inv} = M$.

To test this scheme, we numerically evaluate the required cumulants $K_\pi$ from quantum trajectory simulations. By substituting the third-order cumulants [Fig.~\ref{fig4}(a)] along with the lower-order cumulants into Eqs.~(\ref{eq:M1})--(\ref{eq:M3}), we obtain the reconstructed distribution $\langle M_k \rangle_{\rm inv}$ [Fig.~\ref{fig4}(b)]. A comparison with the statistical average $\langle M_k \rangle$ counted directly from the microscopic trajectories [Fig.~\ref{fig2}(b)] reveals an excellent quantitative agreement. Since Eqs.~(\ref{eq:M1})--(\ref{eq:M3}) are derived entirely from the effective CGF $\tilde{\mathcal F}$, this perfect matching not only verifies that our effective model captures the essential mechanism of the noise dynamics, but also provides a viable experimental protocol to uncover hidden internal distributions.

\emph{Conclusion and Discussions}---In summary, we have established an open quantum dynamical picture for shot noise suppression in dissipative chiral transport. By mapping the system onto a quantum circuit, we reveal that the occupancy distribution and the fluctuation of the total particle number are the fundamental variables determining the noise dynamics. Our proposed effective CGF framework not only unveils the roles of these variables in the noise dynamics, but also provides an experimental protocol to extract them through noise cumulant measurements. Finally, the physical insights and methods developed here go far beyond chiral fermionic systems. We expect our scheme to open an avenue for understanding the shot noise in a much broader class of systems, including diffusive conductors, Majorana excitations in topological superconductors, and fractional quantum Hall edge states.
\let\oldaddcontentsline\addcontentsline
\renewcommand{\addcontentsline}[3]{}

\emph{Acknowledgments}---We are grateful to Alexander Altland for fruitful discussions. This work was supported by KAKENHI Grant No. JP22H01152 from the Japan Society for the Promotion of Science. M. U. was supported by the RIKEN TRIP initiative. M. G. was supported by the China Postdoctoral Science Foundation Grant No. BX20240004. 

\bibliography{draft_sn_ming}

\onecolumngrid 
\vspace{1em}
\begin{center}
	\textbf{End Matter}
\end{center}
\setcounter{equation}{0}
\renewcommand{\theequation}{A\arabic{equation}}
\twocolumngrid 

\emph{Inversion Scheme for the Occupancy Distribution}---To obtain the inverted average occupancy distribution $\langle M_k \rangle_{\rm inv}$ from the measurable joint cumulants $K_\pi = \partial_\pi \tilde{\mathcal{F}}|_{\boldsymbol{\lambda}=0}$, we employ a three-step inversion framework:

\emph{Step 1. Construct the filter cumulants $Q_m$.}

Combine these $K_\pi$ to exactly cancel out all terms containing higher-order correlations (e.g., $\langle M_1 M_2 \rangle$). It can be achieved by constructing the filter cumulants $Q_m$ ($m=1,\dots,N$) as
\begin{equation}
	Q_m = \sum_{\pi \vdash m} C(\pi) K_\pi,
\end{equation}
where the coefficient $C(\pi)$ is determined by the integer partitions $\pi$ of $m$ given by
\begin{equation}
	C(\pi) = (-1)^{|\pi|-1} (|\pi|-1)! \frac{m!}{\prod_{j=1}^{|\pi|} p_j! \prod_{v} c_v!}.
\end{equation}
Here, $|\pi|$ denotes the total number of parts in the partition $\pi$, $p_j$ represents the value of the $j$-th part, and $c_v$ is the multiplicity of the integer $v$ within $\pi$. For example, when $m=2$, there are two partitions: $\pi = (2)$ and $\pi = (1,1)$. For the partition $\pi = (1,1)$, the number of parts is $|\pi|=2$, the components being $p_1=p_2=1$, and the integer $1$ has a multiplicity of $c_1=2$, yielding $C(1,1) = -1$. These $Q_m$ contain only linear combinations of $\langle M_k \rangle$ without any higher-order correlations (see Supplemental Materials for the rigorous proof~\cite{note_supp}), thereby enabling the exact linear inversion written as
\begin{equation}
	Q_m = \sum_{k=1}^N \langle M_k \rangle W_m(k).
\end{equation} 

\emph{Step 2. Calculate the transfer coefficients $W_m(k)$.}

Using set partitions and symmetric polynomials, $W_m(k)$ can be obtained exactly as (see Supplemental Materials for the derivations~\cite{note_supp})
\begin{equation}
	W_m(k) = \sum_{r=1}^m (-1)^{r-1} (r-1)! \begin{Bmatrix} m \\ r \end{Bmatrix} \frac{k^{\underline{r}}}{N^{\underline{r}}},
\end{equation}
where $x^{\underline{r}} := x(x-1)\cdots(x-r+1)$ denotes the falling factorial, and $\begin{Bmatrix} m \\ r \end{Bmatrix}$ denotes the Stirling number of the second kind, which is explicitly given by
\begin{equation}
	\begin{Bmatrix} m \\ r \end{Bmatrix} = \frac{1}{r!} \sum_{j=0}^r (-1)^{r-j} \binom{r}{j} j^m.
\end{equation}

\emph{Step 3. Reconstruct  $\langle M_k\rangle$.}

We can directly formulate the exact inversion as follows:
\begin{equation}
	\begin{pmatrix}
		\langle M_1 \rangle \\
		\langle M_2 \rangle \\
		\vdots \\
		\langle M_N \rangle
	\end{pmatrix}
	=
	\begin{pmatrix}
		W_1(1) & W_1(2) & \cdots & W_1(N) \\
		W_2(1) & W_2(2) & \cdots & W_2(N) \\
		\vdots & \vdots & \ddots & \vdots \\
		W_N(1) & W_N(2) & \cdots & W_N(N)
	\end{pmatrix}^{-1}
	\begin{pmatrix}
		Q_1 \\
		Q_2 \\
		\vdots \\
		Q_N
	\end{pmatrix}.
\end{equation}
It can be proved that the matrix $\mathbf{W}$ is invertible.
\onecolumngrid

\let\addcontentsline\oldaddcontentsline 
\clearpage 
\twocolumngrid
\setcounter{secnumdepth}{3} 
\setcounter{tocdepth}{2}    

\setcounter{equation}{0}
\setcounter{figure}{0}
\setcounter{table}{0}
\setcounter{page}{1}
\setcounter{section}{0}

\makeatletter
\renewcommand{\theequation}{S\arabic{equation}}
\renewcommand{\thefigure}{S\arabic{figure}}
\renewcommand{\thetable}{S\arabic{table}}
\renewcommand{\thepage}{S\arabic{page}} 
\renewcommand{\thesection}{S\arabic{section}}
\makeatother
\onecolumngrid 
\begin{center}
	\textbf{\large Supplemental Materials for ``Open Quantum Theory of Shot Noise in Dissipative Chiral Transport''}
\end{center}
\vspace{3ex}
\twocolumngrid 
\addtocontents{toc}{\protect\setcounter{tocdepth}{2}} 
\tableofcontents
\section{Construction of the Quantum Circuit Model}	\label{Construction of the Quantum Circuit Model}
\subsection{Kraus Operators for Dissipation}	
For a given channel $n$ between adjacent layers $m$ and $m+1$, the explicit forms of these Kraus operators $\hat{K}_\uparrow$, $\hat{K}_\downarrow$, and $\hat{K}_0$ are
\begin{subequations}
	\label{eq:kraus_ops}
	\begin{align}
		\hat{K}_{\uparrow,n}^{(m,m+1)} &= \sqrt{\gamma_\uparrow} \hat{c}_{m+1,n}^\dagger \hat{c}_{m,n}, \quad \hat{K}_{\downarrow,n}^{(m,m+1)} = \sqrt{\gamma_\downarrow} \hat{c}_{m,n}^\dagger \hat{c}_{m+1,n}, \label{eq:kraus_a} \\
		\hat{K}_{0,n}^{(m,m+1)} &= \hat{I} - A_{\uparrow} \hat{P}_{\uparrow,n}^{(m,m+1)} - A_{\downarrow} \hat{P}_{\downarrow,n}^{(m,m+1)}. \label{eq:kraus_b}
	\end{align}
\end{subequations}
Here, $A_{\uparrow/\downarrow}=1 - \sqrt{1 - \gamma_{\uparrow/\downarrow} }$. $\hat{P}_{\uparrow,n}^{(m,m+1)} = \hat{n}_{m,n}(1-\hat{n}_{m+1,n})$ and $\hat{P}_{\downarrow,n}^{(m,m+1)} = \hat{n}_{m+1,n}(1-\hat{n}_{m,n})$ are orthogonal projection operators. By construction, these operators satisfy the completeness relation~\cite{breuer_theory_2007,nielsen_quantum_2010} 
\begin{equation}
	\sum_{\alpha} \hat{K}_{\alpha,n}^{(m,m+1)\dagger} \hat{K}_{\alpha,n}^{(m,m+1)} = \hat{I}
\end{equation}
for any channel $n$ and adjacent layers $(m,m+1)$, with $\alpha \in \{0, \uparrow, \downarrow\}$. The jump parameters satisfy the detailed balance conditions $\gamma_\downarrow=\gamma_0$ and $\gamma_\uparrow = \gamma_0 \exp\left(-\Delta E/T_{\rm bath}\right)$. For a fixed level spacing $\Delta E$, the dissipative process is uniquely determined by $\gamma_0$ and $T_{\rm bath}$.

\subsection{Details of the Quantum Circuit Evolution}	
The time evolution of the brick-wall quantum circuit is discretized into steps, each represents a unit of scattering and dissipation processes. A single step from $\tau-1$ to $\tau$ proceeds in the following four sequential stages: 

(i) Applying jump operators $\hat{\mathbf{K}}_{\{\alpha\}}^{\text{odd}} = \prod_{m \in \text{odd}} \hat{K}_{\alpha_m,1}^{(m,m+1)}$ across odd-even layer pairs; 

(ii) Applying intra-layer random scattering gates $\hat{\mathbf{U}}_{\tau} = \prod_{m} \hat{U}_{\tau}^{(m)}$; 

(iii) Applying jump operators $\hat{\mathbf{K}}_{\{\beta\}}^{\text{even}}$ across even-odd layer pairs;

(iv) Applying a second set of scattering gates $\hat{\mathbf{V}}_{\tau}$. 

The jump operators $\hat{K}_{\alpha_m,1}^{(m,m+1)}$ are restricted to act solely on the first channel for simplicity. This is justified because the subsequent scattering gates $\hat{\mathbf{U}}_{\tau}$ and $\hat{\mathbf{V}}_{\tau}$ mix all channels symmetrically, rendering all channels physically equivalent. The density matrix thus evolves as
\begin{equation}
	\hat\rho(\tau) = \sum_{\{\alpha\},\{\beta\}} \hat{\mathbf{V}}_{\tau} \hat{\mathbf{K}}_{\{\beta\}}^{\text{even}} \hat{\mathbf{U}}_{\tau} \hat{\mathbf{K}}_{\{\alpha\}}^{\text{odd}} \hat\rho(\tau-1) \hat{\mathbf{K}}_{\{\alpha\}}^{\text{odd}\dagger} \hat{\mathbf{U}}_{\tau}^\dagger \hat{\mathbf{K}}_{\{\beta\}}^{\text{even}\dagger} \hat{\mathbf{V}}_{\tau}^\dagger.
	\label{Eq_one_step_CPTP}
\end{equation} 
Equation.~(\ref{Eq_one_step_CPTP}) compactly defines a completely positive trace-preserving (CPTP) superoperator $\mathcal{E}_\tau$. The discrete-time master equation reads $\hat\rho(\tau) = \mathcal{E}_\tau[\hat\rho(\tau-1)]$. Therefore, an $L$-step evolution maps the density matrix to
\begin{equation}
	\hat\rho(L) = \mathcal{E}_L\circ \mathcal{E}_{L-1}\circ \cdots \circ\mathcal{E}_1 [\hat\rho_{\rm in}].
\end{equation}

\subsection{Quantum Trajectory Unraveling}	\label{Quantum Trajectory Unraveling}
We unravel the discrete-time master equation into an ensemble of stochastic quantum trajectories~\cite{breuer_theory_2007,nielsen_quantum_2010}. Each trajectory represents the probabilistic evolution of a pure state. Consider the system from step $\tau-1$ to $\tau$, starting from the state $|\psi(\tau-1)\rangle$. The four-stage evolution yields the intermediate states $|\psi(\tau-1)\rangle_{\rm (i)}$, $|\psi(\tau-1)\rangle_{\rm (ii)}$, $|\psi(\tau-1)\rangle_{\rm (iii)}$, and $|\psi(\tau-1)\rangle_{\rm (iv)} \equiv |\psi(\tau)\rangle$.

Following Eq.~(\ref{Eq_one_step_CPTP}), the unitary scattering operations [stages (ii) and (iv)] are deterministic. The state simply updates as $|\psi(\tau-1)\rangle_{\rm (ii)} = \hat{\mathbf{U}}_\tau |\psi(\tau-1)\rangle_{\rm (i)}$ and $|\psi(\tau-1)\rangle_{\rm (iv)} = \hat{\mathbf{V}}_\tau |\psi(\tau-1)\rangle_{\rm (iii)}$.

The dissipative operations [stages (i) and (iii)] are probabilistic. Focusing on stage (i), we selects a specific jump configuration described by the composite operator $\hat{\mathbf{K}}_{\{\alpha\}}^{\text{odd}}$. The total probability for this multi-layer transition is
\begin{equation}  
	P_{\{\alpha\}} = \langle \psi(\tau-1) | \hat{\mathbf{K}}_{\{\alpha\}}^{\text{odd}\dagger} \hat{\mathbf{K}}_{\{\alpha\}}^{\text{odd}} | \psi(\tau-1) \rangle. 
\end{equation}
Because $\hat{\mathbf{K}}_{\{\alpha\}}^{\text{odd}}$ acts on mutually disjoint layer pairs, this total probability factorizes and can be efficiently evaluated as the product of the individual pair probabilities 
\begin{equation}
	P_{\{\alpha\}} = \prod_{m \in \text{odd}} p_{\alpha_m}^{(m)},
\end{equation}
where 
\begin{equation}
	p_{\alpha_m}^{(m)} = \langle \psi(\tau-1) | \hat{K}_{\alpha_m,1}^{(m,m+1)\dagger} \hat{K}_{\alpha_m,1}^{(m,m+1)} | \psi(\tau-1) \rangle.
\end{equation}
Upon sampling a specific global configuration $\{\tilde{\alpha}\}$ according to $P_{\{\alpha\}}$, the state is updated and renormalized as
\begin{equation}
	|\psi(\tau-1)\rangle_{\rm (i)} = \frac{\hat{\mathbf{K}}_{\{\tilde{\alpha}\}}^{\text{odd}} |\psi(\tau-1)\rangle}{\sqrt{P_{\{\tilde{\alpha}\}}}}.
\end{equation}
Stage (iii) proceeds similarly, updating $|\psi(\tau-1)\rangle_{\rm (ii)}$ to $|\psi(\tau-1)\rangle_{\rm (iii)}$ using $\hat{\mathbf{K}}_{\{\beta\}}^{\text{even}}$.

By generating $N_{\text{traj}}$ independent stochastic trajectories, the density matrix at step $\tau$ is statistically recovered via the ensemble average $\hat\rho(\tau) \approx \frac{1}{N_{\text{traj}}} \sum_{j=1}^{N_{\text{traj}}} |\psi_j(\tau)\rangle \langle \psi_j(\tau)|$. In our implementation, physical observables and noise cumulants are evaluated by directly averaging the expectation values obtained from individual pure-state trajectories, allowing for highly efficient MPS simulations.

\section{Details of the Matrix Product State Implementation}

\subsection{MPS Representation}
\begin{figure}[h]
	\centering
	\includegraphics[width=0.40\textwidth]{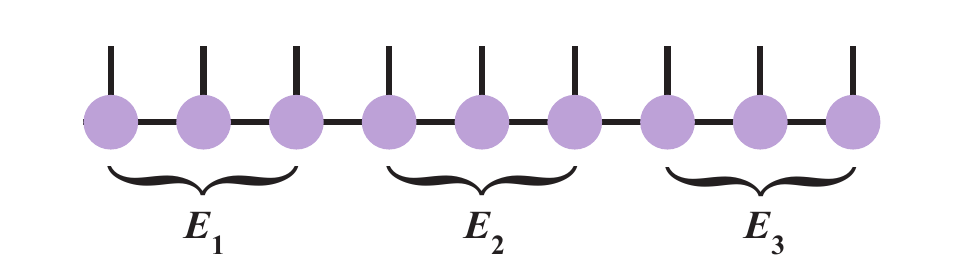}
	\caption{MPS representation of the one-dimensional fermionic chain with three energy levels ($E_1$, $E_2$, and $E_3$) and three channels ($N=3$).}
	\label{S_mps}
\end{figure}
We map the system onto a one-dimensional fermionic chain of length $\mathcal{L} = M \times N$. The site index is denoted as $i = (m-1)N + \alpha$, where $m \in \{1,\dots,M\}$ is the energy level and $\alpha \in \{1,\dots,N\}$ is the channel index, as depicted in Fig.~\ref{S_mps}.
Any pure state $|\psi(\tau)\rangle$ along a quantum trajectory is represented as an MPS~\cite{verstraete_quantum_2009,nielsen_quantum_2010,schollwock_density-matrix_2011,paeckel_time-evolution_2019,weimer_simulation_2021,xiang_density_2023,yadalam_process-tensor_2026,itensor}
\begin{equation}
	|\psi(\tau)\rangle = \sum_{n_1, \dots, n_{\mathcal{L}}} \mathbf{A}^{[1],n_1} \mathbf{A}^{[2],n_2} \cdots \mathbf{A}^{[\mathcal{L}],n_{\mathcal{L}}} |n_1 n_2 \dots n_{\mathcal{L}}\rangle,
\end{equation}
where $n_i \in \{0, 1\}$ is the physical index representing the fermionic occupation at site $i$. Under open boundary conditions, each $\mathbf{A}^{[i],n_i}$ is a matrix of dimension $\chi_{i-1} \times \chi_i$ acting on the virtual bond space. To yield a scalar coefficient, the boundary bond dimensions are strictly fixed to $\chi_0 = \chi_{\mathcal{L}} = 1$, meaning $\mathbf{A}^{[1],n_1}$ is a row vector and $\mathbf{A}^{[\mathcal{L}],n_{\mathcal{L}}}$ is a column vector. The internal bond dimensions are dynamically truncated up to a maximum cutoff $\chi$ (i.e., $\chi_i \le \chi$).

\subsection{Construction of Evolution Operators}
Both the dynamical evolution operators (including the unitary scattering operators and dissipative Kraus operators) and the physical observables are represented by operators applying on MPS. Here, we first delineate the operator constructions for the trajectory dynamics.

\subsubsection{Unitary Scattering}
\begin{figure}[h]
	\centering
	\includegraphics[width=0.39\textwidth]{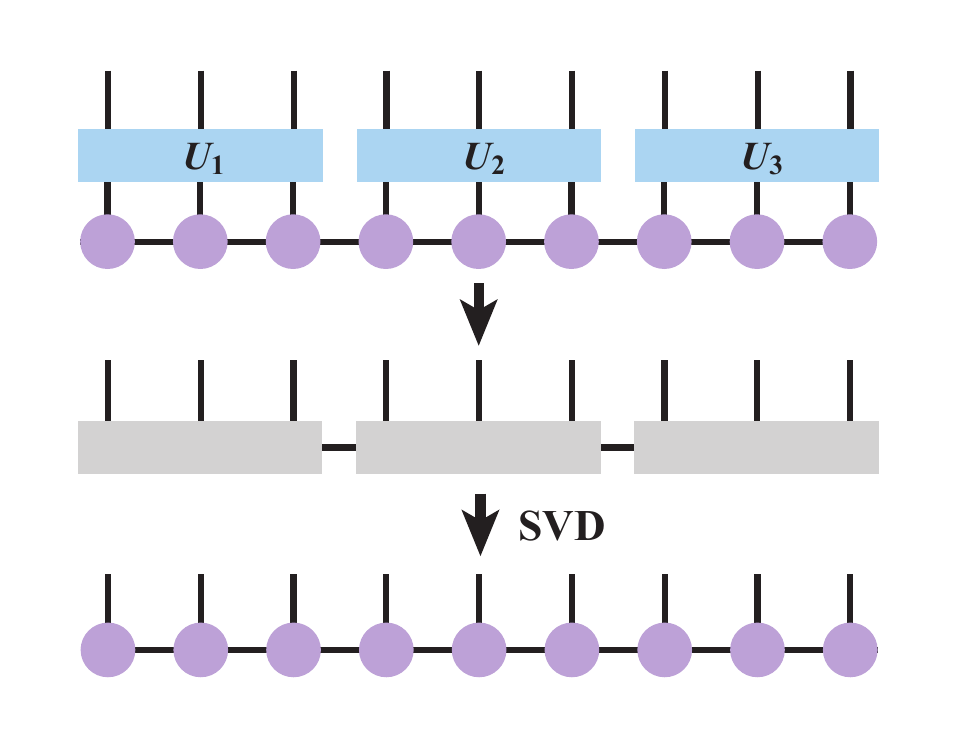}
	\caption{Schematic of the unitary scattering and the subsequent MPS update. The rectangular blocks spanning three adjacent sites (illustrated for $N=3$) represent the application of the Haar-random unitary operators defined in Eq.~(\ref{eq_S_unitary}). SVD is then employed to factorize the multi-site tensors back into the standard $M \times N$ single-site MPS, which simultaneously updates the local tensor elements and the virtual bond dimensions.}
	\label{S_unitary}
\end{figure}
The random scattering at energy level $m$ is characterized by a single-particle Haar-random unitary matrix $\mathbf{s}^{(m)} \in U(N)$. To implement this scattering within the many-body Fock space, we define an effective single-particle Hamiltonian matrix via $\mathbf{H} = i \ln \mathbf{s}^{(m)}$. The corresponding many-body unitary operator is directly given by 
\begin{equation}
	\hat{U}^{(m)} = \exp(-i \sum_{i,j=1}^N H_{ij} \hat{c}_{m,i}^\dagger \hat{c}_{m,j}).
	\label{eq_S_unitary}
\end{equation}
When mapping this operator onto the one-dimensional basis $|n_1 n_2 \dots n_{\mathcal{L}}\rangle$, the fermionic anticommutation relations are rigorously preserved via Jordan-Wigner transformations. The resulting operators and their application are illustrated in Fig. \ref{S_unitary}.

\subsubsection{Dissipative Kraus Operators}
\begin{figure}[h]
	\centering
	\includegraphics[width=0.44\textwidth]{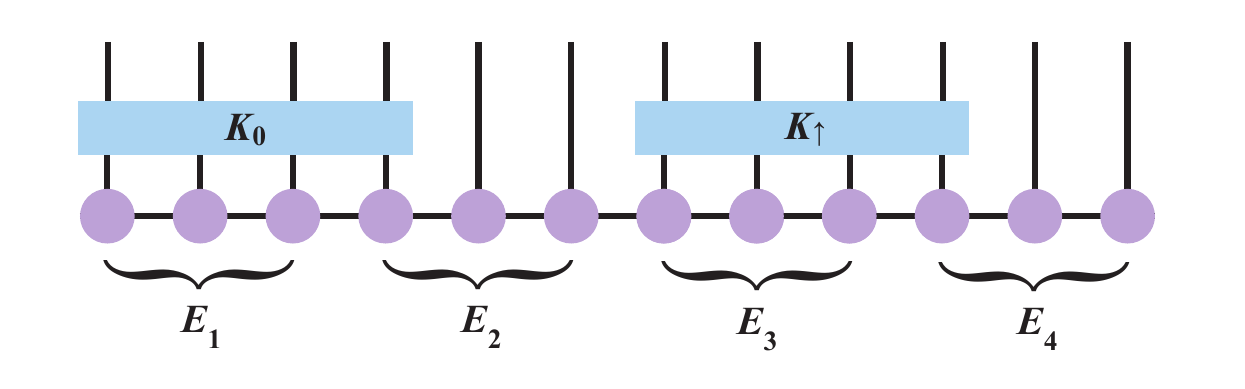}
	\caption{Schematic of the dissipative Kraus operators. The rectangular blocks span all intermediate sites between the first channels of adjacent energy levels. Here, $K_0$ between $E_1$ and $E_2$, and $K_\uparrow$ between $E_3$ and $E_4$ are shown as examples.}
	\label{S_kraus}
\end{figure}
As established in Sec.~\ref{Quantum Trajectory Unraveling}, each step of the dissipative evolution entails applying a specific Kraus operator $\hat{K}$ to the system. Since we consider jumps between the first channels of adjacent energy levels, the corresponding operator spans across all intermediate lattice sites between the two target sites, as schematically illustrated in Fig.~\ref{S_kraus}. The subsequent SVD-based MPS update is similar to that of the unitary scattering case, while the jump probability evaluation follows the exact protocol detailed in Sec.~\ref{Quantum Trajectory Unraveling}.

\subsection{Average of Observables}
\begin{figure}[h]
	\centering
	\includegraphics[width=0.44\textwidth]{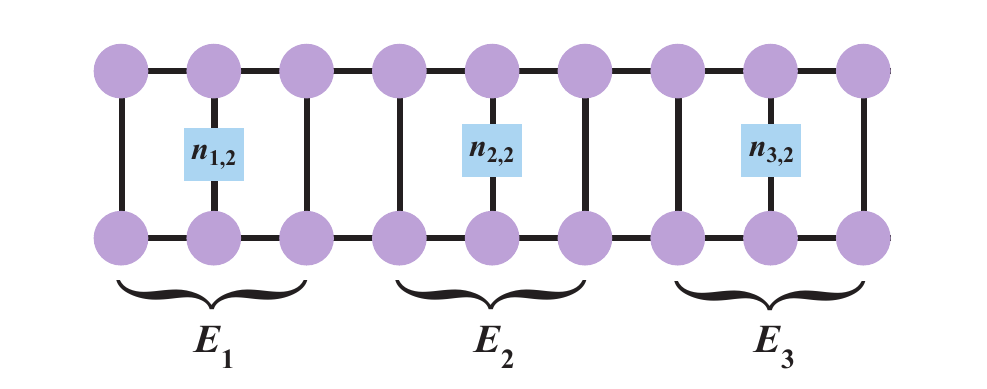}
	\caption{Schematic of the expectation value calculation for local observables. The square blocks represent the local number operators (e.g., $n_{1,2}$, $n_{2,2}$, and $n_{3,2}$) acting on specific channels.}
	\label{S_number}
\end{figure}
Calculating noise cumulants directly requires evaluating the expectations of high-order polynomial operators, such as $\hat{N}_\alpha^2$ and $\hat{N}_\alpha^3$, where $\hat{N}_\alpha = \sum_m \hat{n}_{m,\alpha}$ with $\alpha=1,\dots,N$. We utilize the fermionic idempotency property $\hat{n}_i^2 = \hat{n}_i$ to algebraically expand the operators as
\begin{subequations}
	\begin{align}
		\hat{N}_\alpha^2 &= \sum_{i} \hat{n}_{i,\alpha} + 2 \sum_{i < j} \hat{n}_{i,\alpha} \hat{n}_{j,\alpha}, \\
		\hat{N}_\alpha^3 &= \sum_{i} \hat{n}_{i,\alpha} + 6 \sum_{i < j} \hat{n}_{i,\alpha} \hat{n}_{j,\alpha} + 6 \sum_{i < j < k} \hat{n}_{i,\alpha} \hat{n}_{j,\alpha} \hat{n}_{k,\alpha},
	\end{align}
\end{subequations}
where the summations run over the spatial sites belonging to channel $\alpha$. The operator of one typical term in $\hat{N}_\alpha^3$ is depicted in Fig.~\ref{S_number}. Moreover, cross-correlations like $\hat{N}_1^2 \hat{N}_2$ are similarly expanded by distributing the sum over different channel subsets.

\subsection{Momentum and Cumulant Generating Function}
\begin{figure}[h]
	\centering
	\includegraphics[width=0.36\textwidth]{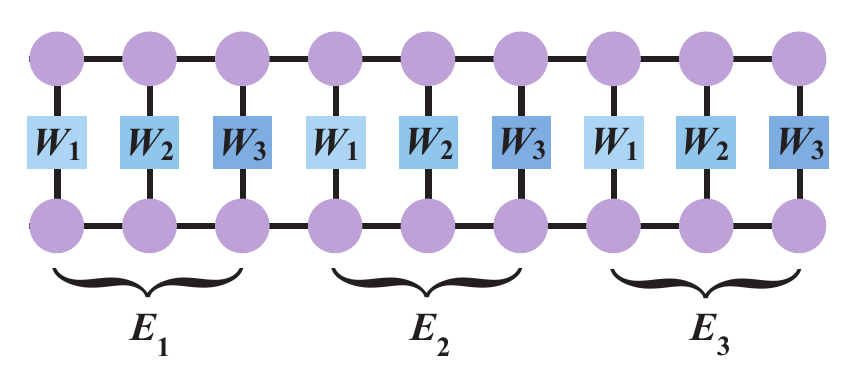}
	\caption{Schematic of the expectation value calculation for the MGF. The square blocks denote the local operators $\hat{W}_{\alpha}=e^{\lambda_\alpha \hat{n}}$ acting on individual channels (illustrated for $\alpha \in \{1, 2, 3\}$), which are repeated across the adjacent energy levels ($E_1$, $E_2$, and $E_3$). The full MGF is evaluated by taking expectation value with respect to the MPS.}
	\label{S_mgf}
\end{figure}
To obtain the CGF, we first evaluate the MGF expressed as
\begin{equation}
	\mathcal{G}(\boldsymbol{\lambda}) = \langle \exp(\sum_{\alpha=1}^N \lambda_\alpha \hat{N}_\alpha) \rangle.
	\label{MGF}
\end{equation} 
Because all local number operators $\hat{n}_i$ commute, the global exponential operator factorizes into the direct product of local operators
\begin{equation}
	e^{\sum_{\alpha} \lambda_\alpha \hat{N}_\alpha} = \prod_{i=1}^{\mathcal{L}} e^{\lambda_{\alpha(i)} \hat{n}_i}.
\end{equation}
Here, $\alpha(i)$ counts the channel indices of site $i$. Exploiting $n_i \in \{0, 1\}$, we have $e^{\lambda \hat{n}_i} = \hat{I} + (e^\lambda - 1)\hat{n}_i$. For each site $i$ we define the local operator $\hat{W}_{\alpha}=e^{\lambda_\alpha \hat{n}}$. Then the expectation value in Eq.~(\ref{MGF}) can be evaluated through this factorized global operator, as depicted in Fig.~\ref{S_mgf}. The CGF is subsequently obtained by taking its logarithm $\mathcal{F}(\boldsymbol{\lambda}) = \ln \mathcal{G}(\boldsymbol{\lambda})$.
\section{Derivation of the Inversion Scheme}
We derive the inversion scheme of $\langle M_k \rangle$ from the effective CGF. Our goal is to reconstruct the average occupancy distribution $\langle M_k \rangle$ ($k=0, \dots, N$). We achieve this by linearly combining the noise cumulants. This step is highly non-trivial since we must find the exact combinations that perfectly cancel out all high-order correlations of ${\mathbf{M}}$.

\subsection{Definitions of the Effective CGF and the Filter Cumulants and Operators}
\subsubsection{The Effective CGF}
Let $\boldsymbol{\lambda} := (\lambda_1, \dots, \lambda_N)$ be the counting field. The effective CGF is defined as
\begin{equation}
	\tilde{\mathcal{F}}(\boldsymbol{\lambda}) = \ln \langle e^{\Psi(\boldsymbol{\lambda})} \rangle_{\mathbf{M}}.
\end{equation}
Here, $\mathbf{M} := (M_0, M_1, \dots, M_N)$. The microscopic generating function is given by
\begin{equation}
	\Psi(\boldsymbol{\lambda}) = \sum_{k=1}^N M_k \ln \left[ S_k(\boldsymbol{\lambda})/\binom{N}{k} \right],
\end{equation}
where $S_k(\boldsymbol{\lambda})=\sum_{i_1 < \dots < i_k} e^{\lambda_{i_1} + \cdots + \lambda_{i_k}}$. The bracket $\langle \cdots \rangle_{\mathbf{M}}$ denotes the statistical average over the configuration ${\mathbf{M}}$.

We expand $\tilde{\mathcal{F}}(\boldsymbol{\lambda})$ into a cumulant series
\begin{equation} 
	\label{eq:cumulant_expansion}
	\tilde{\mathcal{F}}(\boldsymbol{\lambda}) = \langle \Psi \rangle_c  +  \frac{1}{2!} \langle \Psi, \Psi \rangle_c + \frac{1}{3!} \langle \Psi, \Psi, \Psi \rangle_c + \cdots .
\end{equation}
Here, $\langle \cdots \rangle_c$ denotes the cumulant average. The first term contains only the mean values $\langle M_k\rangle$. The higher-order terms inevitably contain high-order correlations of $M_k$. The effects induced by these correlations must be canceled out by systematically combining the joint cumulants of different orders.

\subsubsection{The Filter Cumulant}
We define the $m$-th order filter cumulant $Q_m$ as a linear combination of cumulants
\begin{equation}
	Q_m = \sum_{\pi \vdash m} C(\pi) K_\pi.
\end{equation}
The sum runs over all integer partitions $\pi$ of $m$ with the coefficients
\begin{equation}
	C(\pi) = (-1)^{|\pi|-1} (|\pi|-1)! \frac{m!}{\prod_{j=1}^{|\pi|} p_j! \prod_{v} c_v!}.
\end{equation}
Here, $|\pi|$ denotes the total number of parts in the partition $\pi$, $p_j$ represents the value of the $j$-th part, and $c_v$ is the multiplicity of the integer $v$ within $\pi$. 

Together with the filter cumulant $Q_m$, we can also define the filter operator 
\begin{equation}
	\label{def_Q_op}
	\hat{Q}_m = \sum_{\pi \vdash m} C(\pi) \partial_\pi.
\end{equation}
Specifically, $\hat{Q}_m[\tilde{\mathcal{F}}(\boldsymbol{\lambda})]\Big|_{\boldsymbol{\lambda}=0}=Q_m$. Our final goal is to prove that $\hat{Q}_m$ perfectly eliminates all higher-order correlations of $M_k$ in Eq.~\eqref{eq:cumulant_expansion}, leaving only linear terms
\begin{equation}
	\label{goal}
	\hat{Q}_m[\tilde{\mathcal{F}}(\boldsymbol{\lambda})] \Big|_{\boldsymbol{\lambda}=0} = \sum_{k=1}^N \langle M_k \rangle W_m(k).
\end{equation}
If we find the exact expression for $W_m(k)$, we obtain a set of linearly independent polynomials of degree $m$. We can then build a matrix equation using the measured $\{Q_m\}_{m=1}^N$. Inverting this matrix rigorously yields all $\langle M_k \rangle$.

\subsection{Effect of the Filter Operator}

Before proving Eq.~(\ref{goal}), we first prove a core algebraic property of the operator $\hat{Q}_m$ based on combinatorics: Let $\rho = (\rho_1, \dots, \rho_n)$ be an integer partition. Let $p_\rho(\boldsymbol{\lambda})$ be the corresponding power sum basis in $\boldsymbol{\lambda}$ space
\begin{equation}
	p_\rho(\boldsymbol{\lambda}) = p_{\rho_1}(\boldsymbol{\lambda}) \cdots p_{\rho_n}(\boldsymbol{\lambda}) = \prod_{i=1}^n \left( \sum_{k=1}^N \lambda_k^{\rho_i} \right).
	\label{p_rho}
\end{equation}
When $\hat{Q}_m$ acts on this basis then evaluated at $\boldsymbol{\lambda}=0$, the result depends only on the number of parts $n$
\begin{equation}
	\hat{Q}_m[p_\rho(\boldsymbol{\lambda})]\Big|_{\boldsymbol{\lambda}=0} = m! \delta_{n, 1}.
	\label{prove_Q}
\end{equation}
\noindent
\textit{Proof:}  We prove the above relation in five steps.

\textbf{Step 1: Derivative allocation.} Apply the operator $\partial_\pi = \frac{\partial^m}{\partial \lambda_1^{p_1} \dots \partial \lambda_r^{p_r}}$ to $p_\rho$.  In the expansion of $p_\rho(\boldsymbol{\lambda}) = \prod_{i=1}^n (\sum \lambda_k^{\rho_i})$, each term is a product of $\lambda$ variables. Specifically, the $i$-th bracket, i.e., $(\sum \lambda_k^{\rho_i})$ contributes a power of $\rho_i$ to one chosen variable. To survive the derivative at $\boldsymbol{\lambda}=0$, the expansion must exactly form the monomial $\lambda_1^{p_1} \cdots \lambda_r^{p_r}$. We must therefore assign the $n$ brackets (index set $S=\{1,\dots,n\}$) to the $r$ target variables ($\lambda_1, \dots, \lambda_r$). The sum of the powers $\rho_i$ assigned to $\lambda_j$ must exactly equal $p_j$. 

\textit{Example:} Consider $\rho = (2,1,1)$ and $\partial_\pi = \frac{\partial^4}{\partial \lambda_1^2 \partial \lambda_2^2}$. The basis is $p_\rho = (\sum \lambda_a^2)(\sum \lambda_b^1)(\sum \lambda_c^1)$. To form the required monomial $\lambda_1^2 \lambda_2^2$, we must allocate the first bracket ($\rho_1=2$) to $\lambda_1$. We then allocate both the second and third brackets ($\rho_2=1, \rho_3=1$) to $\lambda_2$. This constitutes one valid allocation. 

Let $M(\rho \to \pi)$ be the total number of valid ordered allocations. After differentiation, each successful match yields the factorial product $\prod_{j=1}^r p_j!$. Thus
\begin{equation} \label{eq:proof_step1}
	\partial_\pi(p_\rho)\Big|_{\boldsymbol{\lambda}=0} = M(\rho \to \pi) \cdot \prod_{j=1}^r p_j!.
\end{equation}

\textbf{Step 2: Factorial cancellation.} Substitute Eq.~\eqref{eq:proof_step1} into the definition of $\hat{Q}_m$ Eq.~(\ref{def_Q_op}). The factorials from the derivative exactly cancel the factorials in the denominator of $C(\pi)$
\begin{equation}
	\hat{Q}_m(p_\rho) \Big|_{\boldsymbol{\lambda}=0} = m! \sum_{\pi \vdash m} (-1)^{r-1} (r-1)! \frac{M(\rho \to \pi)}{\prod c_v!}.
\end{equation}

\textbf{Step 3: Unordered set partitions.} The inner term $\frac{M(\rho \to \pi)}{\prod c_v!}$ divides the number of ordered allocations by the permutation symmetry of the target variables. This operation removes the target labels. It transforms the ordered allocations into unordered set partitions $\Pi$ of the index set $S=\{1, \dots, n\}$.

\textit{Example:} Recall the case $\rho = (2,1,1)$ and $\partial_\pi = \frac{\partial^4}{\partial \lambda_1^2 \partial \lambda_2^2}$. The index set is $S=\{1, 2, 3\}$. We allocated bracket 1 to $\lambda_1$, and brackets 2 and 3 to $\lambda_2$. This is one ordered allocation. Since both $\lambda_1$ and $\lambda_2$ require a squared power ($p_1=2, p_2=2$), they are symmetric targets. Swapping them (brackets 2 and 3 to $\lambda_1$, bracket 1 to $\lambda_2$) creates a second valid allocation. Dividing by the permutation symmetry ($2!$) removes the $\lambda$ labels. Both ordered allocations merge into a single unordered set partition: $\{\{1\}, \{2, 3\}\}$.

Every set partition $\Pi$ maps to a unique integer partition $\pi$. Multiple set partitions can map to the same integer partition. The fraction $\frac{M(\rho \to \pi)}{\prod c_v!}$ exactly counts this multiplicity.  Therefore, summing over $\pi$ is strictly equivalent to summing over all set partitions $\Pi \in \text{SetPartitions}(n)$. The number of target variables $r$ becomes the number of subsets $|\Pi|$
\begin{equation}
	\hat{Q}_m(p_\rho)\Big|_{\boldsymbol{\lambda}=0} = m! \sum_{\Pi \in \text{SetPartitions}(n)} (-1)^{|\Pi|-1} (|\Pi|-1)!.
\end{equation}

\textbf{Step 4: Stirling numbers.} The summand depends only on the number of subsets $k = |\Pi|$. The Stirling number of the second kind, $\begin{Bmatrix} n \\ k \end{Bmatrix}$, counts the exact number of ways to partition $n$ elements into $k$ non-empty subsets. Grouping the summation by $k$ yields
\begin{equation} \label{eq:proof_stirling}
	\hat{Q}_m(p_\rho)\Big|_{\boldsymbol{\lambda}=0} = m! \sum_{k=1}^n \begin{Bmatrix} n \\ k \end{Bmatrix} (-1)^{k-1} (k-1)!
\end{equation}

\textbf{Step 5: Evaluation via power series.} Let $f(n)$ be the sum in Eq.~\eqref{eq:proof_stirling}. We encode $f(n)$ into an exponential power series. Using the known identity for Stirling numbers, 
\begin{equation}
	\sum_{n=k}^\infty \begin{Bmatrix} n \\ k \end{Bmatrix} \frac{t^n}{n!} = \frac{(e^t-1)^k}{k!},
\end{equation}
we evaluate the series as
\begin{align}
	\sum_{n=1}^\infty f(n) \frac{t^n}{n!} 
	&= \sum_{k=1}^\infty (-1)^{k-1} (k-1)! \left( \sum_{n=k}^\infty \begin{Bmatrix} n \\ k \end{Bmatrix} \frac{t^n}{n!} \right) \nonumber \\
	&= \sum_{k=1}^\infty (-1)^{k-1} (k-1)! [ (e^t - 1)^k/k! ] \nonumber \\
	&= \ln\left(1 + (e^t - 1)\right) \nonumber \\
	&= t
\end{align}
Comparing the coefficients of $t^n$, we find $f(1)=1$. For all $n \ge 2$, $f(n)=0$. Substituting this back into Eq.~\eqref{eq:proof_stirling} completes the proof. \hfill$\blacksquare$

As a direct consequence of Eq.~(\ref{prove_Q}), we have the following corollary: The operator $Q_m$ perfectly annihilates all higher-order correlations of $M_k$ in Eq.~\eqref{eq:cumulant_expansion}.

\textit{Proof:} Recall the generating function $\Psi(\boldsymbol{\lambda}) = \sum_{k=1}^N M_k L_k(\boldsymbol{\lambda})$, where $L_k = \ln \left[ S_k(\boldsymbol{\lambda})/\binom{N}{k} \right]$. The function $L_k$ is symmetric in $\boldsymbol{\lambda}$ and vanishes at $\boldsymbol{\lambda}=0$. By the fundamental theorem of symmetric polynomials and Newton's identities, $L_k$ expands entirely in terms of power sum polynomials $p_{\rho_i}=\sum_{k=1}^N \lambda_k^{\rho_i}$. Recall that a basis $p_\rho$ for a partition $\rho = (\rho_1, \cdots, \rho_n)$ is a product of individual power sums. We write the Taylor expansion of $L_k$ explicitly as
\begin{equation}
	L_k = \sum_{m=1}^\infty \sum_{\rho \vdash m} c_\rho\cdot p_{\rho_1} p_{\rho_2} \cdots p_{\rho_n},
\end{equation}
where $\sum_{i}\rho_i=m$. The second summation runs over all $\rho = (\rho_1, \cdots, \rho_n)$ that satisfy this condition. Because $L_k$ has no constant term, every product basis in this expansion contains at least one factor. Now consider any $l$-th order cumulant ($l \ge 2$), such as $\langle \Psi, \Psi \rangle_c$. It generates terms like $M_{k_1} M_{k_2} \cdots M_{k_l}$, associated with the product of $l$ functions $L_{k_1} L_{k_2} \dots L_{k_l}$. Expanding this product means multiplying their power sum bases. The expanded product exclusively contains new bases with a total part count $n \ge l \ge 2$. Through Eq.~(\ref{prove_Q}), the operator $\hat{Q}_m$ strictly annihilates any basis with $n \ge 2$. Therefore, only the linear term ($l=1$) survives
\begin{equation}
	\label{Q_ann}
	\hat{Q}_m[\tilde{\mathcal{F}}(\boldsymbol{\lambda})]\Big|_{\boldsymbol{\lambda}=0} = \hat{Q}_m[\langle \Psi \rangle_c]\Big|_{\boldsymbol{\lambda}=0} = \sum_{k=1}^N \langle M_k \rangle W_m(k),
\end{equation}
with $W_m(k) = \hat{Q}_m [ \ln( S_k(\boldsymbol{\lambda})/\binom{N}{k} ) ]\Big|_{\boldsymbol{\lambda}=0}$. \hfill$\blacksquare$

\subsection{Determining the Coefficients $W_m(k)$}
We derive the exact expression of $W_m(k)$ in three steps.

\textbf{Step 1: Combinatorial expansion.} Let $A$ be a subset of $\{1, \dots, N\}$ with size $|A|=k$. We write
\begin{equation}
	S_k(\boldsymbol{\lambda}) = \sum_{|A|=k} \prod_{i \in A} e^{\lambda_i}.
\end{equation}
We substitute $x_i = e^{\lambda_i} - 1$. The equation becomes
\begin{equation} \label{eq:Sk_combo}
	S_k(\boldsymbol{\lambda})= \sum_{|A|=k} \prod_{i \in A} (1 + x_i).
\end{equation}
Expand the product $\prod_{i \in A} (1+x_i)$. This means choosing a sub-subset $B$ of size $r$ ($0 \le r \le k$) from $A$. Swapping the summation order and fix a product of $r$ variables $\prod_{j \in B} x_j$. This specific term appears exactly $\binom{N-r}{k-r}$ times across all possible subsets $A$. Grouping these terms gives
\begin{equation} \label{eq:Sk_expanded}
	S_k(\boldsymbol{\lambda}) = \sum_{r=0}^k \binom{N-r}{k-r} \sum_{|B|=r} \prod_{j \in B} x_j.
\end{equation}
We define the inner sum as the $r$-th elementary symmetric polynomial $e_r(\mathbf{x})$.  It explicitly sums all possible products of $r$ distinct variables
\begin{equation}
	e_r(\mathbf{x}) \equiv \sum_{1 \le j_1 < \dots < j_r \le N} x_{j_1} x_{j_2} \cdots x_{j_r}.
\end{equation}
Now we construct the normalized ratio present in our target logarithmic function. Dividing Eq.~\eqref{eq:Sk_expanded} by $\binom{N}{k}$ yields
\begin{equation}
	S_k(\boldsymbol{\lambda})/\binom{N}{k} = \sum_{r=0}^k \mu_r e_r(\mathbf{x}),
\end{equation}
where we define the coefficient $\mu_r$. Using falling factorials $x^{\underline{r}} := x(x-1)\cdots(x-r+1)$, it simplifies to
\begin{equation}
	\mu_r \equiv \binom{N-r}{k-r}/\binom{N}{k} = \frac{k^{\underline{r}}}{N^{\underline{r}}}.
\end{equation}
We isolate the $r=0$ term. Since $\mu_0=1$ and $e_0(\mathbf{x})=1$, this zeroth-order term is exactly 1. We define the remaining sum ($r \ge 1$) as a new polynomial $Y(\boldsymbol{\lambda})$
\begin{equation}
	Y(\boldsymbol{\lambda}) \equiv \sum_{r=1}^k \mu_r e_r(\mathbf{x}).
\end{equation}
Consequently, the normalized partition function is $1 + Y(\boldsymbol{\lambda})$. Then $L_k(\boldsymbol{\lambda})$ can be expressed as
\begin{equation}
	\label{Lk}
	L_k(\boldsymbol{\lambda}) = \ln\left[ 1 + Y(\boldsymbol{\lambda}) \right].
\end{equation}

\textbf{Step 2: Logarithmic linearization.} Expanding Eq.~(\ref{Lk}) and applying $\hat{Q}_m$ yields
\begin{equation}
	W_m(k) = \hat{Q}_m \left[ Y - \frac{1}{2}Y^2 + \frac{1}{3}Y^3 - \cdots \right]\Big|_{\boldsymbol{\lambda}=0}.
\end{equation}
The function $Y(\boldsymbol{\lambda})$ is strictly symmetric in $\boldsymbol{\lambda}$, and vanishes at the origin ($Y(\mathbf{0})=0$). By similar reason in proving Eq.~(\ref{Q_ann}), $\hat{Q}_m$ eliminates all terms including multiple $Y$. Then we have
\begin{equation} 
	\label{eq:Wm_linearized}
	W_m(k) = \hat{Q}_m [Y]\Big|_{\boldsymbol{\lambda}=0} = \sum_{r=1}^N \frac{k^{\underline{r}}}{N^{\underline{r}}} \hat{Q}_m [ e_r(\mathbf{x}) ]\Big|_{\boldsymbol{\lambda}=0}.
\end{equation}

\textbf{Step 3: Final expression.} Now we evaluate $\hat{Q}_m [ e_r(\mathbf{x}) ]\Big|_{\boldsymbol{\lambda}=0}$. The unique matching term yields $\partial_\pi [ e_r(\mathbf{x}) ]\Big|_{\boldsymbol{\lambda}=0} = \delta_{|\pi|, r}$.
Applying $\hat{Q}_m$ and sum the coefficients $C(\pi)$ yields
\begin{equation}
	\hat{Q}_m [ e_r(\mathbf{x}) ]\Big|_{\boldsymbol{\lambda}=0} = \sum_{\pi \vdash m, |\pi|=r} C(\pi) = (-1)^{r-1}(r-1)! \begin{Bmatrix} m \\ r \end{Bmatrix}.
\end{equation}
Substitute this into Eq.~\eqref{eq:Wm_linearized}. We obtain
\begin{equation}
	W_m(k) = \sum_{r=1}^m (-1)^{r-1} (r-1)! \begin{Bmatrix} m \\ r \end{Bmatrix} \frac{k^{\underline{r}}}{N^{\underline{r}}}.
\end{equation}

Algebraically, $W_m(k)$ is a polynomial in $k$ of degree $m$. Evaluating the degree-increasing basis $\{W_m(k)\}_{m=1}^N$ at $N$ distinct points ($k=1, \dots, N$) generates a Vandermonde-type matrix, leading to non-zero determinant. This guarantees invertibility of $\mathbf{W}$, ensuring a unique exact solution for $\langle M_k \rangle$.

\section{Simulation Parameters}

Figures.~2 and~4 share the same set of model parameters (in the following we set $\Delta E=1$):
\begin{itemize}
	\setlength{\itemsep}{1.5pt}    
	\setlength{\parsep}{1.5pt}     
	\setlength{\parskip}{1.5pt}    
	\item \textbf{System Size:} Total number of energy levels $L = 19$, with $N = 3$, totally $N_{\rm site}=57$ sites.
	
	\item \textbf{Chemical Potentials:} $\mu_1 = 18.1$ for the injection channel 1, and $\mu_{2} = \mu_{3} = 0.1$ for the remaining channels.
	
	\item \textbf{Temperatures:} Incident temperature $T_{\text{in}} = 10^{-5}$ for all channels; bath temperature $T_{\text{bath}} = 10^{-6}$.
	
	\item \textbf{Dissipation Coupling:} Base jump rate $\gamma_0 = 0.7$.
	
	\item \textbf{Simulation Evolution:} Total steps $= 120$. 
	
	\item \textbf{Statistical Averaging:} Observables are averaged over an ensemble of $N_{\text{dis}} = 100$ random disorder configurations (random $\hat{\mathbf{U}}$ configurations), with $N_{\text{traj}} = 100$ quantum trajectories sampled each configuration.
\end{itemize}

Parameters for Fig.~3:
\begin{itemize}
	\setlength{\itemsep}{1.5pt}    
	\setlength{\parsep}{1.5pt}     
	\setlength{\parskip}{1.5pt}    
	\item \textbf{System Size:} $L = 25$, with $N = 3$, resulting in a total of $N_{\rm site}=75$ sites.
	
	\item \textbf{Chemical Potentials:} $\mu_1 = 17.1$, and $\mu_{2} = \mu_{3} = 8.1$. Note that $\mu_{2}$ and $\mu_{3}$ define the reference ground chemical potential. (We write $\mu_{\rm in}=9 \Delta E$ in the main text for this reason.) We set this  finite value rather than zero to properly resolve the Fermi-Dirac distribution at finite temperatures, preventing unphysical truncation of the thermal tails.
	
	\item \textbf{Temperatures:} The incident temperature $T_{\text{in}}$ (applied identically to all three channels) is scanned from $10^{-5}$ (effectively $0$) to $4.0$. The bath temperature $T_{\text{bath}}$ is scanned from $10^{-5}$ to $1.0$. Both parameter ranges are uniformly discretized into $14$ points.
	
	\item \textbf{Dissipation Coupling:} Base jump rate $\gamma_0 = 0.99$.
	
	\item \textbf{Simulation Evolution:} Total steps $= 70$. 
	
	\item \textbf{Statistical Averaging:} Observables are averaged over an ensemble of $N_{\text{dis}} = 20$, with $N_{\text{traj}} = 40$ quantum trajectories sampled each configuration.
\end{itemize}

\end{document}